\documentclass[aps,prl,amsmath,amssymb,superscriptaddress,showkeys,showpacs,twocolumn,floatfix]{revtex4-1}

\usepackage{graphicx}
\usepackage{subfigure}
\usepackage{hyperref}
\usepackage{ulem}

\usepackage{xcolor}
\usepackage[utf8]{inputenc}

\begin{document}



\title{Cosmology meets functional QCD: First-order cosmic QCD transition induced by large lepton asymmetries}

\author{Fei Gao}
\email[]{f.gao@thphys.uni-heidelberg.de}
\affiliation{Institut f{\"u}r Theoretische Physik,
	Universit{\"a}t Heidelberg, Philosophenweg 16,
	69120 Heidelberg, Germany}
\author{Isabel M. Oldengott}
\email[]{isabel.oldengott@uclouvain.be}
\affiliation{Centre for Cosmology, Particle Physics and Phenomenology, Universit\'e catholique de Louvain, Louvain-la-Neuve B-1348, Belgium}
\affiliation{Departament de Fisica Te\`orica and IFIC, CSIC-Universitat de Val\`encia, 46100 Burjassot, Spain}

\date{\today}

\begin{abstract}
The lepton flavour asymmetries of the Universe are observationally almost unconstrained before the onset of neutrino oscillations. We calculate the cosmic trajectory during the cosmic QCD epoch in the presence of large lepton flavour asymmetries.
By including QCD thermodynamic quantities derived from functional QCD methods in our calculation our work reveals for the first time the possibility of a first-order cosmic QCD transition. We specify the required values of the lepton flavour asymmetries for which a first-order transition occurs for a number of different benchmark scenarios.
\end{abstract}


\maketitle

\paragraph{Introduction}

According to the hot big bang model, the early Universe has experienced (at least) two epochs where phase transitions could occur: the electroweak transition at temperatures around $T_{ew}\sim 100$ GeV when fermions and gauge bosons became massive particles;  and the transition of quantum chromo dynamics (QCD) at $T_{QCD}\sim 130$ MeV when quarks confined into hadrons.

The Standard Model (SM) of particle physics predicts both the electroweak as well as the QCD transition to be crossovers. However,  various  observations from cosmology and particle physics unavoidably hint towards the existence of physics beyond the SM.
Many extensions of the SM would render the electroweak transition to be first-order (for the most simple extension see \cite{Espinosa:1993bs}), while there are much fewer mechanisms known that could provide a first-order QCD transition (\cite{Iso:2017uuu,Hambye:2018qjv}). %
A first-order QCD transition would have an impact on the spectrum of gravitational waves and the production of exotic relics \cite{Schwarz:2003du,Witten:1984rs}.
The details of the QCD transition furthermore have an impact on the production of primordial black holes \cite{Jedamzik:1996mr,Jedamzik:1999am,Byrnes:2018clq,Vovchenko:2020crk,Bodeker:2020stj} which are one of the prime candidates for dark matter.

In this work, we show for the first time that large lepton flavour asymmetries can render the QCD transition to be first-order. The fact that the value of the lepton asymmetry has an impact on the cosmic QCD epoch has already been known for a while and been discussed in a series of papers \cite{Schwarz:2009ii,Wygas:2018otj,Middeldorf-Wygas:2020glx}. However, so far it has not been possible to provide a definite answer to the question whether a sufficiently large lepton asymmetry can indeed  induce a first-order transition and what \textit{sufficiently large}  means quantitatively.

\paragraph*{QCD diagram}
QCD matter experiences a transition from quarks and gluons to hadrons as temperature and chemical potential changes.
To chart the phase diagram of QCD for non-zero temperature, charge- and baryon chemical potential has attracted a lot of interest.
Lattice QCD simulations have confirmed a crossover at vanishing chemical potential \cite{Karsch:2001vs, Aoki:2006we, Borsanyi:2010cj,Bazavov:2014pvz}. However, due to the sign problem, it is difficult for lattice QCD to explore the region of large chemical potentials, leaving the possibility of  the existence of a critical end point (CEP) and a region of a first-order transition unsettled. On the observational front, searching for signals of the QCD phase transition, especially the CEP, are the main goals of the current and the future experimental program at the relativistic heavy ion collider \cite{Luo:2017faz, Adamczyk:2017iwn, Andronic:2017pug, Shuryak:2014zxa}. On the theoretical front, large chemical potentials are accessible by continuum QCD methods, i.e. functional QCD methods including Dyson-Schwinger equations (DSEs) \cite{Binosi:2009qm,Roberts:2000aa,Eichmann:2016yit, Fischer:2018sdj} and the functional renormalization group (fRG) method \cite{Pawlowski:2005xe,Dupuis:2020fhh}.
The functional QCD methods are non-perturbative continuum methods which are capable of describing  both the dynamical chiral symmetry
breaking (DCSB) and the confinement simultaneously. Though limited by truncations, the functional QCD approach has made fruitful progress in studying the QCD phase structure and thermal properties~\cite{Qin:2010nq,Fischer:2018sdj,Fu:2019hdw,Gao:2020qsj} in a large range of chemical potential.

Concerning the cosmic QCD epoch, the question about the nature of the QCD transition is actually two-fold: The first part of the question is if QCD matter under any circumstances could ever experience a first-order phase transition. Even though different functional QCD methods all predict the existence of a CEP (with varying locations)~\cite{Fischer:2018sdj,Dupuis:2020fhh} its final confirmation is still awaiting experimental evidence. The second part is under which conditions such a first-order transition would have been realized in the early Universe. This work deals with the second part of the question by applying results from functional QCD.

\paragraph{Large Lepton Asymmetry}
When the Universe expands and cools down it follows a certain path in the QCD phase diagram, the so called \textit{cosmic trajectory}. The standard cosmic trajectory is based on the assumption that matter and antimatter are (almost) equally abundant in the Universe. Indeed, measurements of primordial element abundances and the anisotropy spectrum of the Cosmic Microwave Background (CMB) show that the Universe only has a tiny asymmetry in the baryonic sector, characterized by the \textit{baryon asymmetry} $b=(8.7 \pm 0.06)\times 10^{-11}$ (inferred from \cite{Aghanim:2018eyx} and where $b$ is going to be properly defined in eq. \eqref{eq:b}).  Explaining the creation of this tiny number necessitates theories of baryogenesis and leptogenesis. The idea of leptogenesis is to create an asymmetry in the leptonic sector and transfer this lepton asymmetry into a baryon asymmetry by sphaleron processes. Therefore, the lepton asymmetry would be of the same order of magnitude as the baryon asymmetry, or more explicitly $l=-\frac{51}{28} b$ \cite{Kolb:1983ni}. In this case the cosmic trajectory passes the QCD diagram at (extremely) small chemical potential where lattice calculations reveal a crossover. Observational constraints from big bang nucleosynthesis (BBN) \cite{Pitrou:2018cgg} and the CMB \cite{Oldengott:2017tzj} on the value of the lepton asymmetry however allow values of the lepton asymmetry as large as $|l|<1.2 \times10^{-2}$ \cite{Oldengott:2017tzj}, i.e. around $8-9$ orders of magnitude larger than the baryon asymmetry. From a theoretical point of view, there are also alternative models predicting a large lepton asymmetry \cite{Canetti:2012kh,Drewes:2021nqr,Eijima:2017anv,Ghiglieri:2018wbs,Harvey:1981cu,Affleck:1984fy}. In any case, lepton asymmetry is a key parameter to understand the origin of the matter-antimatter asymmetry of our Universe.

Observationally even less constrained than the \textit{total} lepton asymmetry $l \equiv  l_e+l_{\mu}+l_{\tau}$ are the individual lepton flavour asymmetries $l_{\alpha=e,\mu,\tau}$: It has been shown \cite{Wong:2002fa,Dolgov:2002ab} that initially different lepton flavour asymmetries are equilibrated at around $T_{osc} \sim 10$ MeV when neutrino oscillations become efficient such that finally $l_{\alpha} \approx \frac{l}{3}$ \footnote{Note however that depending on the initial values and signs of $l_\alpha$ and depending on the mixing angles equilibration may however only be partial \cite{Pastor:2008ti,Barenboim:2017dfq,Johns:2016enc}}. This implies that CMB and BBN constraints are only sensitive to $l$ but have no constraining power on the initial values of the individual lepton flavour asymmetries $l_{\alpha}$. From an agnostic point of view, the individual lepton flavour asymmetries should therefore be treated as free parameters (while their sum still needs to fulfill the observational constraint $|l|<1.2 \times 10^{-2}$ \cite{Oldengott:2017tzj}). As recently pointed out in \cite{Mukaida:2021sgv}, large lepton flavour asymmetries could also explain the baryon asymmetry of the Universe.

\paragraph{Cosmic trajectory}

Let us in the following explain how we calculate the cosmic trajectory. The basic idea of the method is similar to the one outlined in \cite{Schwarz:2009ii,Wygas:2018otj,Middeldorf-Wygas:2020glx}.

Within the framework of the SM, before the onset of neutrino oscillations at $T_{osc}\approx 10$ MeV and after the electroweak transition at $T_{ew}\approx 100$ GeV electric charge, baryon number and lepton flavours are conserved in a comoving volume.  Assuming as well entropy conservation in comoving volume, the baryon $b$, lepton flavour number $l_{\alpha}$ and electric charge $q$  asymmetry can be written as

\begin{eqnarray}
\label{eq:l}
l_{\alpha}&&=\frac{n_{\alpha}+n_{\nu_{\alpha}}}{s} \hspace{0.5cm} (\alpha = e,\mu,\tau) , \\
\label{eq:b}
b&&=\sum_i \frac{B_i n_i}{s} , \\
\label{eq:q}
q&&=\sum_i \frac{Q_i n_i}{s} ,
\end{eqnarray}

\noindent and \textit{remain constant} throughout the evolution of the Universe,  for temperatures $T_{ew}>T>T_{osc}$. The sum in eq. \eqref{eq:b} goes over all baryons and in eq. \eqref{eq:q} over all charged particles. $n_i$ denotes the net number density (i.e. particle minus anti-particle number density) of particle species $i$ and $s$ denotes the total entropy density. As mentioned above, the baryon asymmetry is a known quantity and we fix it to the central value $b=8.7\times 10^{-11}$ from \cite{Aghanim:2018eyx}. There is furthermore good reason to believe that the Universe is charge neutral \cite{Caprini:2003gz} such that we may set $q=0$ from now on.  While the standard assumption is $|l_{\alpha}|= \mathcal{O}(b)$, we treat the three lepton flavour asymmetries $l_{\alpha}$ as free input parameters within observational constraints. This  may be realized by some new physics violating lepton flavour around some temperature $T_{BSM}$ well before the QCD epoch, such that the $l_{\alpha}$ in eq. \eqref{eq:l} are conserved at $T_{BSM}>T>T_{osc}$. We assume electroweak processes to maintain kinetic equilibrium, i.e. all particles sharing the same temperature $T_i=T$, and chemical equilibrium which allows us to find relations between the chemical potentials of different particle species, e.g. $\mu_u=\mu_c$ etc. Note that we assume no other but the particle content of the Standard Model throughout this work.
Each conserved charge is associated with a chemical potential (i. e. $\mu_{\text{B}}$, $\mu_{\text{Q}}$ and $\mu_{L_{\alpha}}$), which can be related to the chemical potentials of the different particle species by (see also e.g. \cite{Schwarz:2009ii} or \cite{Kapusta:2006pm})
\begin{eqnarray}
\mu_{L_{\alpha}} &=& \mu_{\nu_{\alpha}}, \label{eq:mu_L} \\
\mu_Q &=& \mu_{u}-\mu_{d}, \label{eq:mu_Q} \\
\mu_B &=& \mu_u + 2 \mu_d \label{eq:mu_B}.
\end{eqnarray}

The cosmic trajectory denotes the solution of eqs. $\eqref{eq:l}-\eqref{eq:q}$ for $(\mu_{L_{\alpha}},\mu_B,\mu_Q)$ at different temperatures $T$ and given a choice for the lepton  asymmetries $l_{\alpha}$.
 As first noted in \cite{Schwarz:2009ii}, large lepton asymmetries $l_{\alpha}$ do not only shift the cosmic trajectory towards large leptochemical potentials $\mu_{L_{\alpha}}$ but also towards large charge- and baryon chemical potentials.

In practice, for leptons and photons we assume Fermi-Dirac or respectively Bose-Einstein distributions for the phase-space densities, in accordance with the assumption of thermal and chemical equilibrium. Concerning the DCSB and confinement phenomena,  modelling the QCD sector at temperatures close to the QCD transition is however much more complicated.

Let us briefly summarize the approach of \cite{Wygas:2018otj,Middeldorf-Wygas:2020glx} before we explain how we treat  QCD matter in this work. In \cite{Wygas:2018otj,Middeldorf-Wygas:2020glx} the QCD sector was treated separately in three different (but overlapping) temperature regimes: i) at high $T$ quarks and gluons were treated as free particles,  where \cite{Middeldorf-Wygas:2020glx} also included contributions from perturbative QCD, ii) at $T \sim T_{QCD}$ the QCD pressure was Taylor expanded and susceptibilities from lattice QCD were  applied, iii) at low $T$ QCD matter was described by the hadron resonance gas (HRG) model. As pointed out in \cite{Middeldorf-Wygas:2020glx}, the method  has two limitations. The appearance of a Bose-Einstein condensate of pions for large lepton asymmetries (see also \cite{Vovchenko:2020crk}) would require a separate treatment of the low-energy modes of pions. The more stringent restriction  however comes from the application of the Taylor expansion of the QCD pressure which is only justified for relatively small chemical potentials.

In this work, we therefore follow a different approach and consult results from functional QCD. In particular, we apply the thermodynamic quantities derived from DSEs in the rainbow-ladder(RL) truncation as it is currently the \textit{only QCD based method that has delivered a complete computation of the phase diagram and the related thermal properties} \cite{Gao:2015kea,Isserstedt:2019pgx,Fischer:2018sdj}. The obtained thermal quantities are qualitatively and partly quantitatively in accordance with those from Lattice QCD simulations ~\cite{Fischer:2018sdj} and can be consistently extended to large chemical potentials.
The existence of a CEP is apparent as a sudden drop in the number density $n_{u/d}$ of the u/d quarks below a certain threshold for the temperature $T_{\text{CEP}}=125$ MeV and above a certain threshold for the chemical potential $\mu_{\text{CEP}}=111$ MeV (figure provided in the appendix \ref{appendix}). 
Let us however note here that rather than being able to determine the exact location of the CEP the strength of the RL truncation is to capture the main properties of the QCD phase structure. Therefore, our work should also be understood as a \textit{proof-of-principle of the  possibility of a first-order cosmic QCD} induced by large lepton flavour asymmetries -- and not as a calculation enabling the exact determination of cosmic trajectories during the QCD epoch.

\paragraph*{Results}
The computational details of our method are summarized in the appendix which includes
refs.~\cite{Isserstedt:2019pgx,Gao:2015kea,Isserstedt:2020qll,Philipsen:2012nu,Gao:2020fbl,Schwarz:2009ii,Wygas:2018otj,Middeldorf-Wygas:2020glx}.
Throughout this work, we focus on the choices for the lepton asymmetries  $l_{\alpha}$ summarized in tab. \ref{tab:l_values}.

As noted before, case (i) is constrained by CMB observations to values $|l|<1.2\times10^{-2}$ \cite{Oldengott:2017tzj}. Cases (ii)-(vii) were chosen such that the total lepton asymmetry $l$ always fulfills the CMB constraints. Note that there is of course an infinite number of realizations for scenarios with $l_e \neq l_{\mu} \neq l_{\tau}$ satisfying the CMB and BBN constraints. Scenarios (ii)-(vii) simply serve as benchmark models.

The main question we are interested in is whether large lepton flavour asymmetries can indeed render the cosmic QCD transition first-order, as it has been speculated in \cite{Schwarz:2009ii,Wygas:2018otj,Middeldorf-Wygas:2020glx}. This requires to work out if at $T=T_{\text{CEP}}$ we find either $\mu_u \geq \mu_{\text{CEP}}$ or $\mu_d \geq \mu_{\text{CEP}}$ (or both). The s and c quark can generally also experience a first-order transition but since their CEP is expected to be at larger chemical potentials we expect this to happen for larger $l_{\alpha}$. We therefore concentrate on the $u$ and $d$ quark in the following. To quantify for which value of $l_{\alpha}$ a first-order transition may happen we add
\begin{equation}
\mu_{i} \geq 111 \, \text{MeV} \hspace{1cm} \text{at} \, \, \, T=125 \, \text{MeV}
\label{eq:CEP_condition}
\end{equation}
as an additional equation on top of eqs. \eqref{eq:l}-\eqref{eq:q}, where $i$ is equal to \textit{either} $u$ \textit{or} $d$.
 All cases (i)-(vii) considered in this work are characterized by only one degree of freedom which is  eliminated by the additional constraint in eq. \eqref{eq:CEP_condition}, i.e. no further input is required in order to solve eqs. \eqref{eq:l}-\eqref{eq:q}.

\begin{table}
\begin{tabular}{|c|c|c|c|}
\hline
& & $\mu_u \geq 111$ MeV & $\mu_d \geq 111$ MeV \\
\hline
(i)& $l_{e}=l_{\mu}=l_{\tau}=\frac{l}{3}$ & $l \geq 1.10 \times 10^{-1}$ & $l\leq -1.03 \times 10^{-1}$ \\
\hline
(ii) & $l_{e}=0,l_{\mu}=-l_{\tau}$ & $l_{\mu} \geq 7.43 \times 10^{-2}$ & $l_{\mu} \leq -6.85\times 10^{-2}$ \\
\hline
(iii) & $l_{e}=-l_{\tau},l_{\mu}=0$ & $l_e \geq 7.14 \times 10^{-2}$ & $l_e \leq -6.59\times 10^{-2}$ \\
\hline
(iv) & $l_{e}=-l_{\mu},l_{\tau}=0$ & $l_e \geq 1.36 \times 10^{-3}$ & no solution \\
\hline
(v) & $l_e=l_{\mu},l_{\tau}=-2 l_{e}$ & $l_e \geq 3.46\times 10^{-2}$ & $l_e \leq -3.23\times 10^{-2}$ \\
\hline
(vi)& $l_e=l_{\tau},l_{\mu}=-2 l_{e}$ & $l_e \leq -1.20 \times 10^{-1}$ & $l_e \geq 1.02\times 10^{-1}$ \\
\hline
(vii) & $l_e=-2l_{\mu},l_{\mu}=l_{\tau}$ & $l_e \leq -1.14\times 10^{-1}$ & $l_e \geq 9.43\times 10^{-2}$ \\
\hline
\end{tabular}
\caption{Values of the lepton flavour asymmetry for which either the u (left column) or the d quark (right column) experiences a first-order transition.}
\label{tab:l_values}
\end{table}

The first result is the fact that there exist solutions to the set of equations, i.e. for large enough lepton asymmetries \textit{the Universe experienced a first-order QCD transition}.
Table \ref{tab:l_values} summarizes for which values of the lepton flavour asymmetries either the u or the d quark experiences a first-order phase transition. For the equilibrated case (i) we see that $l$ has to be larger than allowed by observations of the CMB \cite{Oldengott:2017tzj} and primordial element abundances \cite{Pitrou:2018cgg}. We therefore conclude that in case of equal lepton flavour asymmetries a first-order QCD transition is  excluded. On the other hand, scenarios (ii)-(vii) are by construction always compatible with CMB and BBN constraints and  all allow a first-order transition.
It is also interesting to compare tab. \ref{tab:l_values} to the findings of \cite{Middeldorf-Wygas:2020glx} where scenarios (i) and (ii) were studied and to \cite{Vovchenko:2020crk} where a more general version of scenario (v) was investigated ($|l_e+l_{\mu}|=-|l_{\tau}|$). The values for $l_{\alpha}$ rendering a first-order transition in tab. \ref{tab:l_values} are remarkably close to the maximal values ensuring the reliability of the Taylor expansion in \cite{Middeldorf-Wygas:2020glx}, namely $|l_{\mu}| \leq 4\times 10^{-2}$ (case (ii)) and $|l| \leq 7.5\times 10^{-2}$ (case (i)). Furthermore, for scenarios (i) and (ii) the values in tab. \ref{tab:l_values} exceed the values for pion condensation, i.e. $|l|\geq 9\times 10^{-2}$ (case (i)) and $|l_{\mu}|\geq 6\times 10^{-2}$ (case (ii)) \cite{Middeldorf-Wygas:2020glx}. For scenario (v) they are only slightly smaller than what is needed to potentially form a pion condensate, i.e.  $|l_{e}+l_{\mu}|>1 \times 10^{-1}$ \cite{Vovchenko:2020crk}. We therefore conclude that in case of a first-order cosmic QCD transition induced by large lepton asymmetries the formation of a pion condensate is also likely.

\begin{figure}
\includegraphics[width=0.98\columnwidth]{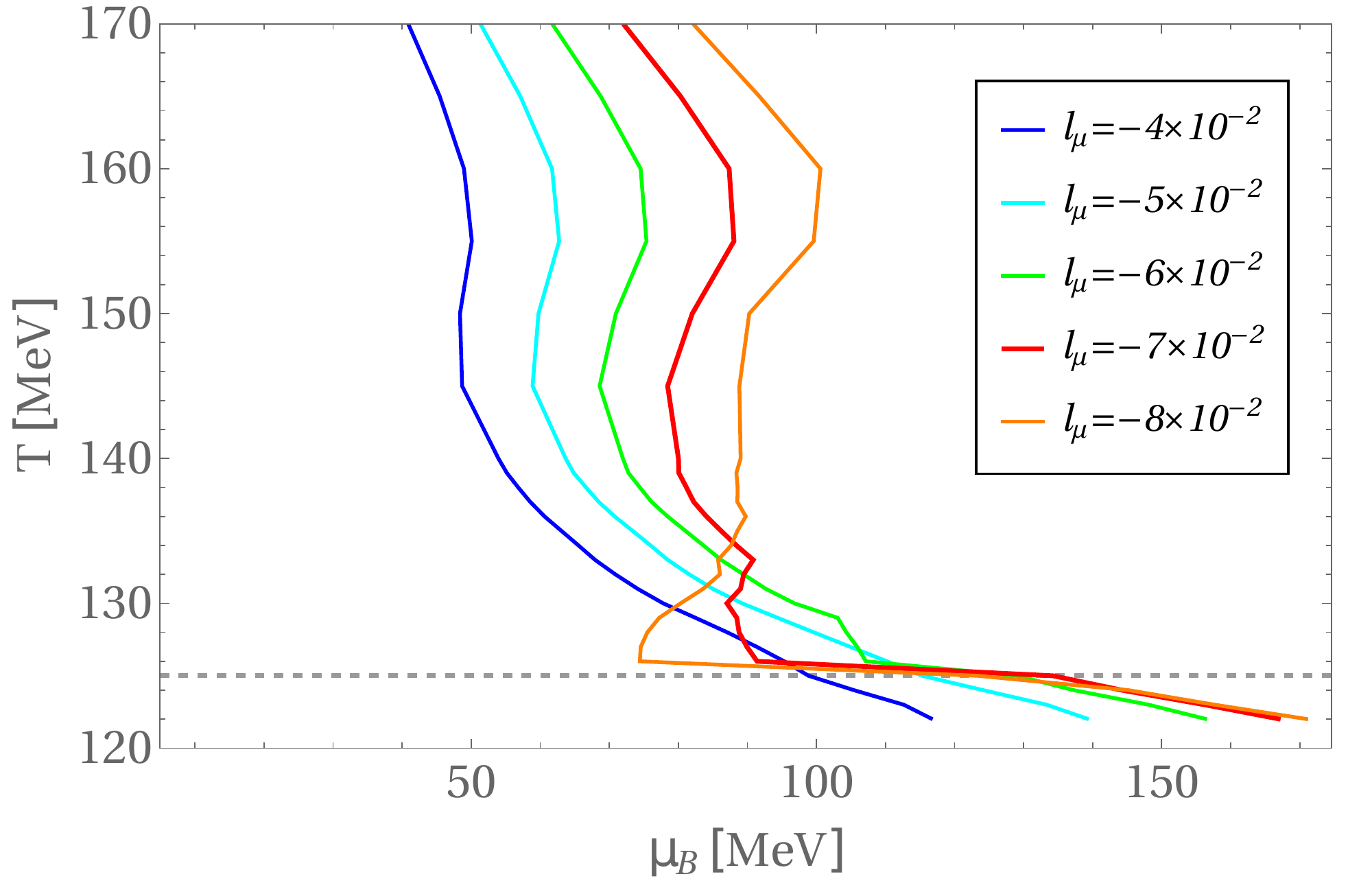}
\includegraphics[width=0.98\columnwidth]{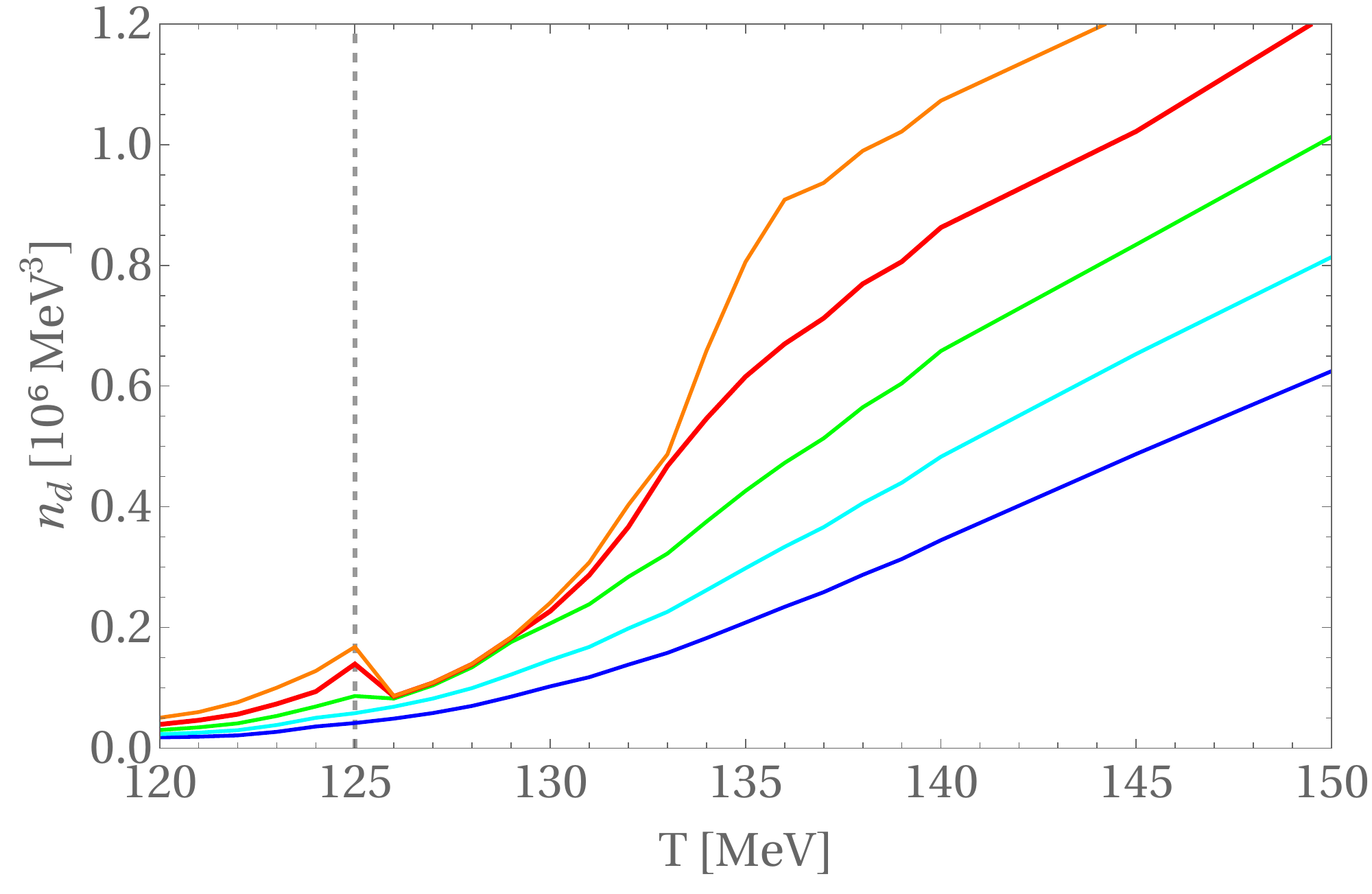}
\caption{Cosmic trajectory (top) and d quark number density (bottom) for scenario (ii), traversing a first-order QCD transition for increasing values of $|l_{\mu}|$. The dashed line marks $T_{\text{CEP}}$.}
\label{fig:trajectory_firstOrder}
\end{figure}

It is furthermore interesting to see how the cosmic trajectory looks like for such a first-order transition. Exemplary we thereby focus on scenario (ii) and a first-order transition of the d quark. Fig. \ref{fig:trajectory_firstOrder} shows the cosmic trajectory projected onto the $(\mu_B,T)$-plane for different values of the lepton flavour asymmetry $l_{\mu}$. Observe the appearance of a kink in the trajectory for increasing values of $|l_{\mu}|$. This discontinuous feature reflects a first-order transition and arises for $l_{\mu}$ values between the green ($l_{\mu}=-6 \times 10^{-2}$) and the red curve ($l_{\mu}=-7\times 10^{-2}$), confirming thereby the solution $l_{\mu} \leq - 6.85 \times 10^{-2}$ given in tab. \ref{tab:l_values}. However, on top of the jump at $T_{\text{CEP}}=125$ MeV the red and orange (first-order) trajectories also show some wiggly features at temperatures above $T_{\text{CEP}}$. This (unphysical) behaviour is caused by cumulative errors appearing at large chemical potential in the computation of the entropy density with the RL truncation (see appendix).
When studying the number density of the d quark, the first-order signal is however much better pronounced (lower plot in fig. \ref{fig:trajectory_firstOrder}), appearing as a sudden jump at $T=125$ MeV.

\begin{figure}[t]
\includegraphics[width=\columnwidth]{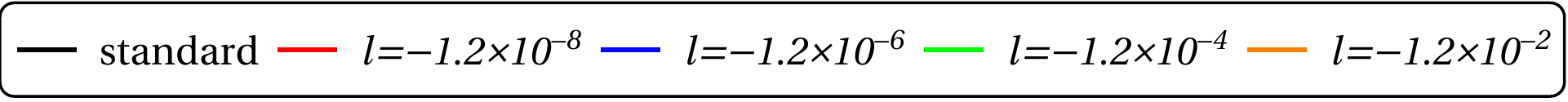}
\includegraphics[width=\columnwidth]{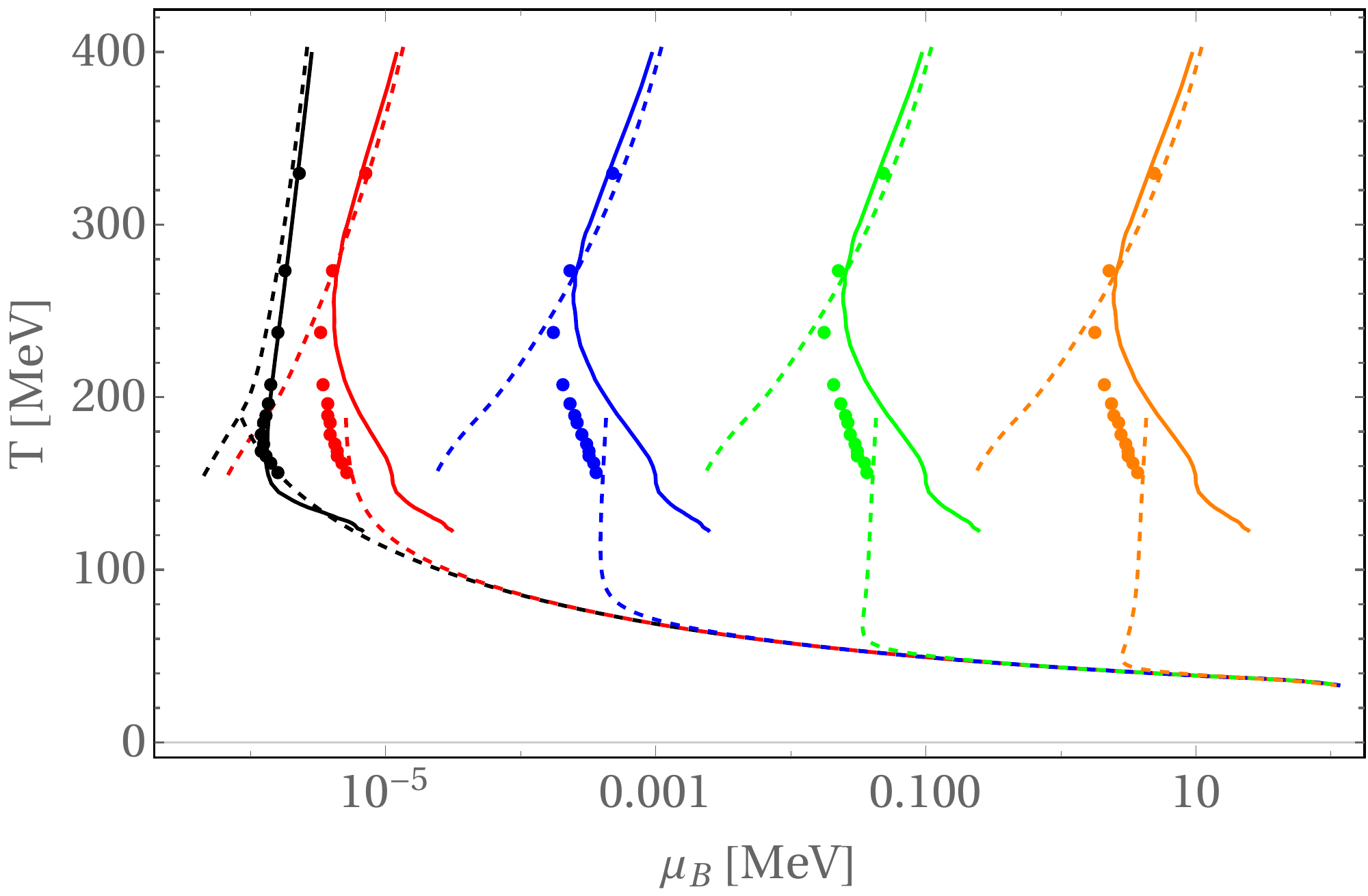}
\caption{Cosmic trajectory projected onto the $(\mu_B,T)$-plane for equal lepton flavour asymmetries (case (i)) for different values of $l$, calculated with the method described in this work (solid lines) and with the method of \cite{Wygas:2018otj,Middeldorf-Wygas:2020glx} (dashed lines denoting high- and low-temperature regimes, $\bullet$ using lattice susceptibilities). Black line assumes $l=-\frac{51}{28}b$. }
\label{fig:trajectory_comparison}
\end{figure}

Finally, let us compare the method introduced in this work to the method of \cite{Wygas:2018otj,Middeldorf-Wygas:2020glx} for the case of equal lepton flavour asymmetries (i). In fig. \ref{fig:trajectory_comparison} we show the cosmic trajectory derived by the method introduced in this work (solid lines) and by the method described in \cite{Middeldorf-Wygas:2020glx}. We see that at high temperatures our new method produces trajectories that are relatively similar to the ones obtained by \cite{Wygas:2018otj,Middeldorf-Wygas:2020glx}. In the intermediate temperature regime the solid lines show the same trend as the dashed lines (i.e. the curves bending in the same direction) but the bending of the solid lines is much stronger. At low temperatures the standard trajectory converges to the HRG line relatively smoothly. However, for all larger $l$ values instead of converging to the dashed HRG lines the solid lines rather overshoot. This discrepancy is supposedly due to the fact that the RL truncation only allows to capture the main properties of the QCD phase structure but is not expected to deliver exact results, as stated above. In the same manner also the solutions in tab. \ref{tab:l_values} are not expected to be exact.

The truncation scheme can be systematically improved and in particular in \cite{Fu:2019hdw, Gao:2020fbl,Gunkel:2021oya} the CEP was estimated as $(T_{\text{CEP}},\mu_{\text{CEP}})_{u/d}\simeq(110,200)$ MeV.
The derivation of the thermal quantities with the method of \cite{Gao:2020fbl} is however computationally more costly than the determination of the CEP and currently still in progress. We plan to incorporate the results into our method in a future study and expect  the agreement to the method of \cite{Wygas:2018otj,Middeldorf-Wygas:2020glx} in fig. \ref{fig:trajectory_comparison} to be improved. 

\paragraph{Conclusions}
In this work, we included the thermodynamic quantities of QCD matter derived from DSEs into calculations of the cosmic trajectory during the QCD epoch. 
Bearing in mind the limitations of the applied RL truncation, our method  offers an up-to-now unique possibility to study the cosmic trajectory in a consistent way over a wide range of temperature and chemical potential and covering the main properties of QCD matter. In particular, for the first time our work revealed that large lepton flavour asymmetries can induce a first-order cosmic QCD transition. For equal lepton flavour asymmetries $l_{\alpha}$ this would require a total lepton asymmetry $l$ that is already ruled out by observations of the CMB and BBN. For unequal lepton flavour asymmetries a first-order transition is however feasible. Examples thereof are given by the cases (ii)-(vii) studied in this work and the corresponding values in tab. \ref{tab:l_values}.

\begin{acknowledgments}
\paragraph{Acknowledgments}
We thank Dominik J. Schwarz, Dietrich B\"odeker, Jan M. Pawlowski and J\"urgen Schaffner-Bielich for interesting discussions and valuable comments. F. G. also thanks Dominik J. Schwarz and Dietrich B\"odeker for invitation and hosting in Bielefeld University.
I. M. O. acknowledges support by Fonds de la recherche scientifique (FRS-FNRS), as well as from the  FPA2017-845438 and the Generalitat Valenciana under grant PROMETEOII/2017/033, and by the Deutsche
Forschungsgemeinschaft (DFG) through the Grant No. CRC-TR 211. F. G. is supported by Alexander von Humboldt foundation.
\end{acknowledgments}

\bibliography{Literature}

\section*{Appendix}
\label{appendix}

\paragraph*{Computation of thermal quantities} The required thermal quantities, number density and entropy density,  can be computed after solving the quark gap equation of DSEs~\cite{Isserstedt:2019pgx,Gao:2015kea,Isserstedt:2020qll}. We here apply the  same settings for the gap equation as described in Ref.~\cite{Gao:2015kea}, and then  solve it through Broyden's iteration method in Fortran.

The number density for the u/d, s and c quarks can be directly computed from the solution of the quark gap equation after subtraction, i.e. through the quark propagator $G$,
\begin{equation}
\label{eq:numb}
n(T,\mu)=-\int \frac{d^3\vec{p}}{(2\pi)^3}\big\{T\sum-\int\frac{dq_0}{2\pi}\big\}({\rm tr}[\gamma_0 G])\, .
\end{equation}
The entropy density can then be expressed as the integral along the chemical potential as
\begin{equation}
\label{eq:pressure}
\delta s(T,\mu)=s(T,\mu)-s(T,\mu=0)=\int_0^\mu \frac{\partial n(T,\mu')}{\partial T} d\mu'.
\end{equation}

Due to the lack of gluon pressure in the approach of Ref.~\cite{Gao:2015kea}, we here  use the  lattice QCD computation at  zero chemical potential with $N_f=2+1+1$ as parametrized  in Ref.~\cite{Philipsen:2012nu}. The full entropy is  given as:
\begin{equation}
s_{\rm QCD}=s_{\rm latt}(T,\mu=0)+\delta s(T,\mu).
\end{equation}

\begin{figure}
\includegraphics[width=0.9\columnwidth]{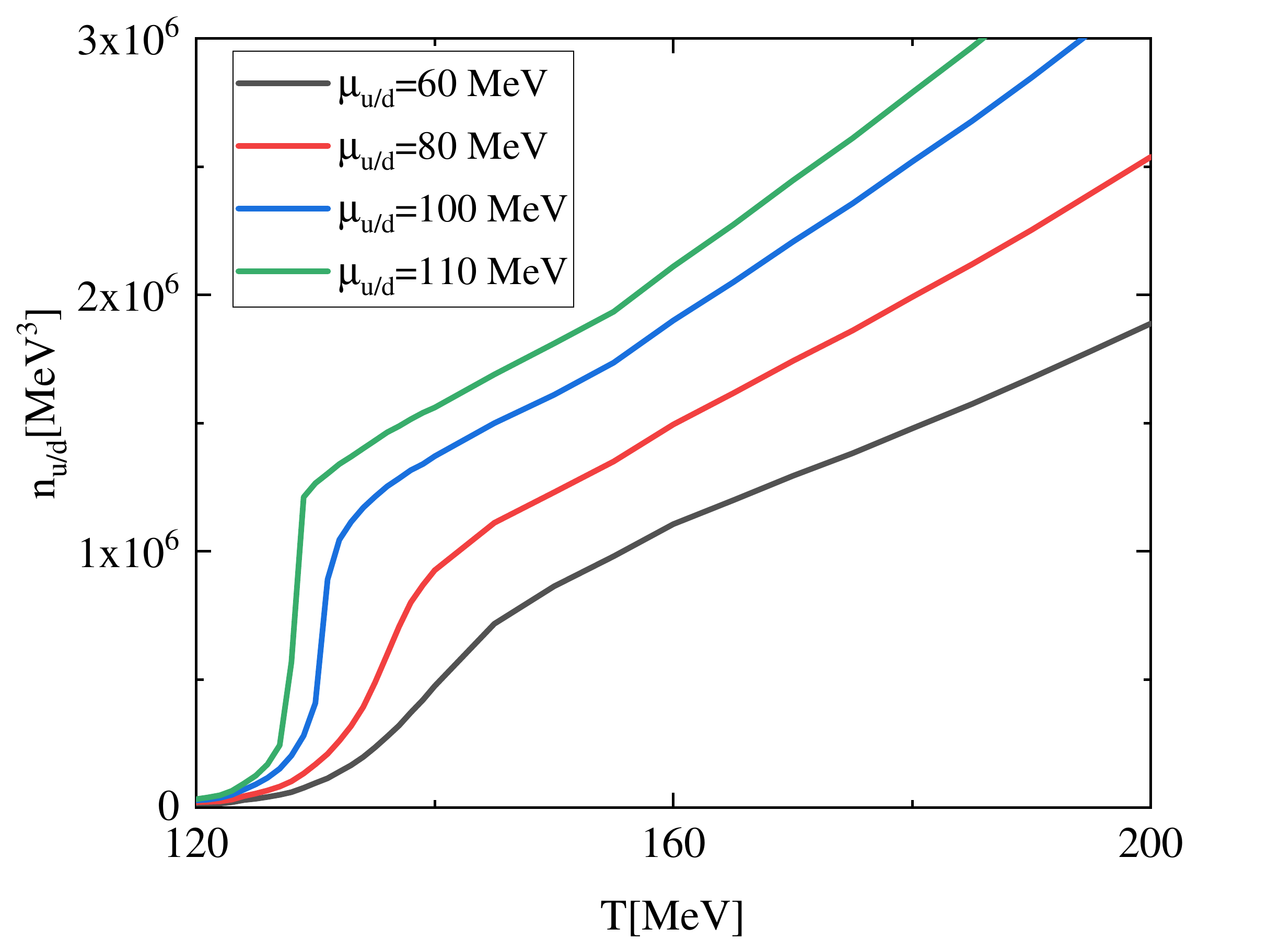}
\includegraphics[width=0.9\columnwidth]{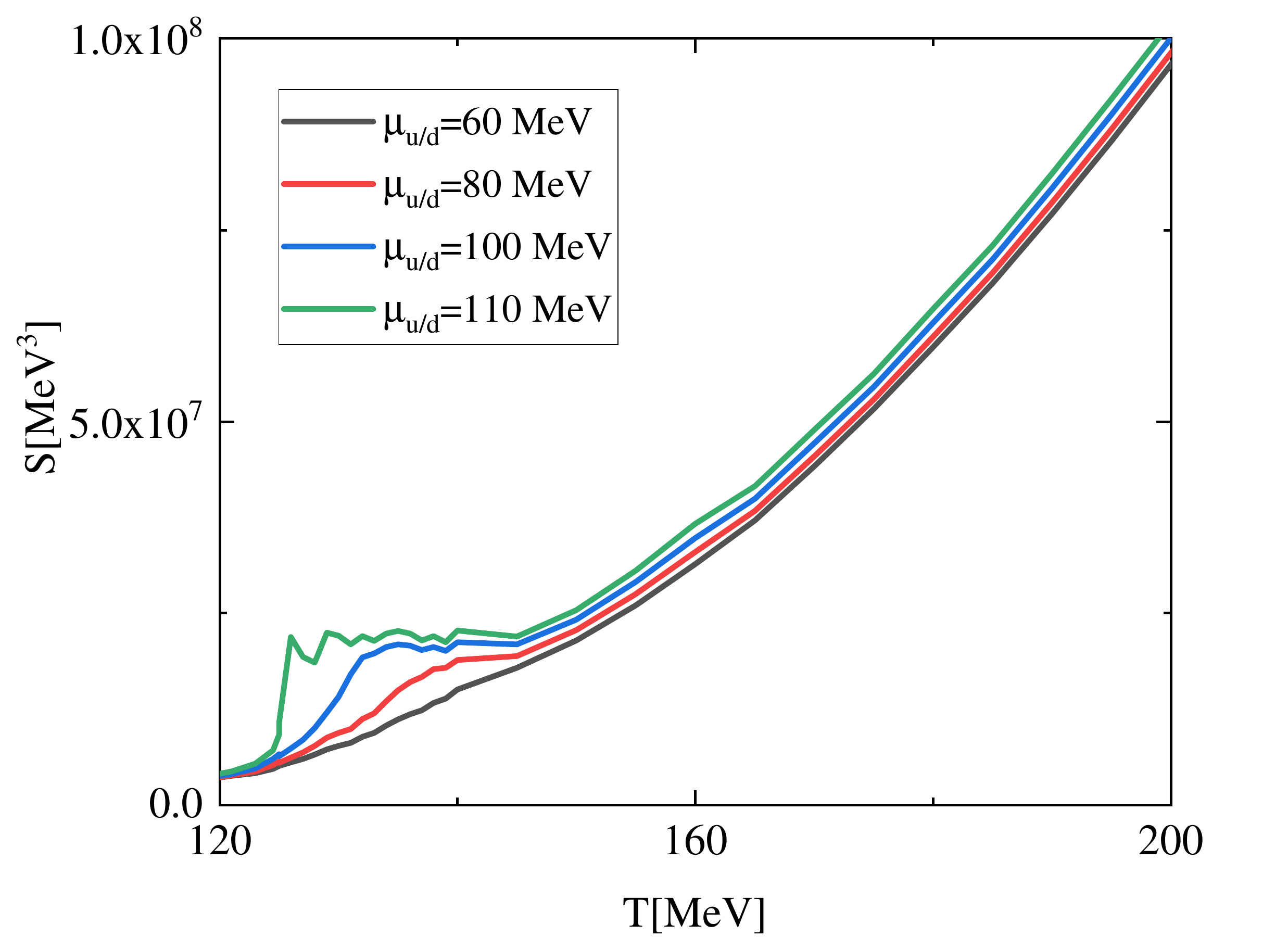}
\caption{Number density (top) and the entropy density (bottom) of the u/d quark as a function of $T$ for different chemical potentials $\mu_{u/d}$. }
\label{fig:ThermalQuantity}
\end{figure}

 The data of the number density and the entropy density are generated on a grid of values for temperature and chemical potential.  The grid spans over the temperature range $T=121-400$ MeV, where we chose a grid size of $1$ MeV at $T<140$ MeV, a grid size of $5$ MeV a $140<T<300$ MeV and a sparse grid size of $20$ MeV at $300<T<400$ MeV. Even though the data for $T<121$ MeV are available, the truncation has been much less reliable in this regime. The number density below $T<120$ MeV gives an odd rising behaviour as $T$ decreases, which can possibly be solved by improving the truncation. Nevertheless, the temperature range we chose still covers the CEP and first-order phase transition region.

The pion condensation effect on the quark propagator is not included in the current truncation. Such an effect is  in the hadron resonance channel of the quark gluon vertex which is the subleading contribution~\cite{Gao:2020fbl}, and hence  will barely have an impact on the QCD transition.  In some certain cases, the pion condensation phase transition and QCD transition can take place successively. In $\mu$-direction the grid spans over the range $0<\mu<1000$ MeV, with a uniform grid size of $1$ MeV. Both number densities and entropy densities of the different particle species were stored in data files.

The data for the number density are generally smooth, except for the characteristic jump when the CEP is crossed, see fig.  \ref{fig:ThermalQuantity} (top). However, almost negligible (unphysical) changes in the curvature of the number density get amplified significantly once the derivative in $T$ is taken in order to compute the entropy density according to eq. \eqref{eq:pressure}. This issue becomes increasingly severe for large chemical potentials since the integral in eq. \eqref{eq:pressure} cumulates small errors in the $\mu$-dependence of the number density. Therefore -- on top of the distinctive plateau that is expected for a first-order transition -- the entropy density shows a wiggly behaviour for large chemical potentials, see fig. \ref{fig:ThermalQuantity} (bottom). This unphysical feature of the entropy density causes the cosmic trajectories to look rather wiggly too for large lepton asymmetries.

\paragraph*{Computation of the cosmic trajectory}
In order to calculate the cosmic trajectory we modified the c-code developed in \cite{Schwarz:2009ii,Wygas:2018otj,Middeldorf-Wygas:2020glx} such that it reads in those data files and interpolates them. The code uses Broyden's method in order to find solutions of eqs. (1)-(3) in the main text for $(\mu_{L_e},\mu_{L_{\mu}},\mu_{L_{\tau}},\mu_B,\mu_Q)$, given the above specified temperature values and with $l_{\alpha}$ as input parameters. In order to  interpolate the data for number and entropy density in $\mu$-direction we applied cubic spline interpolation.

\end{document}